# De Newton a Einstein: el nacimiento de la Relatividad Especial[*]

*Rafael Ferraro*

Instituto de Astronomía y Física del Espacio, y
Departamento de Física, FCEyN, Universidad de Buenos Aires

Frecuentemente Mario viaja desde Rosario a Buenos Aires, y ha verificado con su propio reloj que el viaje tiene una duración de 3hs. Hoy Mario comprará un regalo en uno de los comercios de la estación de Buenos Aires. Para evitar inconvenientes, Mario decide reservar telefónicamente su compra. El tren está a punto de partir, cuando Mario telefonea al empleado del comercio porteño. Mario observa las 14.00 en el reloj de la estación de Rosario, que opera en perfecta sincronización con los relojes de las demás estaciones. Por lo tanto, le comunica al empleado del comercio que retirará su compra a las 17.00. Ya son las 17.00; el tren ingresa en la estación y el empleado aguarda a Mario con el regalo listo para ser retirado. Este modo usual de organizar encuentros en nuestra vida cotidiana está basado en la tácita aceptación de que el tiempo transcurre de igual manera en distintos "laboratorios" en movimiento relativo. Las tres horas de Mario a bordo del tren son también tres horas para los relojes sincronizados del sistema ferroviario en tierra.

Aunque Mario no lo sabía, ese día un equipo de empiristas se propuso confirmar que tampoco las longitudes son alteradas por el movimiento. Con cierta perplejidad, Mario había observado que su vagón llevaba adherida una cinta métrica, que atestiguaba la longitud de 50m. En tanto los empiristas, provistos de una cinta métrica idéntica que habían desplegado junto a la vía, aguardaron el paso del tren con la intención de medir la longitud del vagón en movimiento. El delicado experimento requería la utilización de sensores capaces de dar la posición de un extremo del vagón en el mismo instante en que el otro extremo pasaba por el cero de la cinta métrica. El previsible resultado fue que la longitud del vagón en movimiento era de 50m. Conformes con la verificación experimental del carácter *invariante* de la

---

longitud (es decir, independiente del laboratorio), los empiristas se reunieron para redactar un artículo donde comunicarían el resultado de su trabajo a la comunidad científica internacional.

Los relatos precedentes muestran que nuestra relación cotidiana con el espacio y el tiempo parece enseñarnos que el tiempo transcurrido entre dos eventos es independiente del estado de movimiento del laboratorio, y que la longitud de un cuerpo es, asimismo, independiente del estado de movimiento del cuerpo. Este aparente carácter invariante de distancias y tiempos constituye uno de los aspectos esenciales de nuestra percepción ingenua del espacio y el tiempo.

El carácter invariante de distancias y tiempos no es la única propiedad "evidente" atribuida al espacio y al tiempo. También aceptamos que el espacio satisface los postulados de la geometría de Euclides, la misma geometría *plana* que se enseña en la escuela, y la única geometría desarrollada hasta mediados del siglo XIX. De los axiomas y postulados de Euclides se obtienen conocidos resultados, como el teorema de Pitágoras o que los ángulos internos de un triángulo suman 180°. Con respecto al tiempo, consideramos que su fluir es uniforme.

Nuestras creencias acerca de la naturaleza del espacio y el tiempo parecen estar tan sólidamente demostradas en nuestra vida cotidiana, que resulta completamente natural que las mismas hayan sido incorporadas en la construcción newtoniana que dominó el pensamiento científico hasta comienzos del siglo XX. En los *Principia* Isaac Newton (1642-1727) presenta al espacio absoluto como el escenario inmutable donde suceden los fenómenos físicos. En el espacio absoluto, la distancia entre dos eventos es una propiedad de los lugares absolutos donde los eventos suceden. La geometría del espacio absoluto es euclidiana (la única concebible en aquella época); por lo tanto el espacio admite sistemas de coordenadas cartesianas, cuyo carácter rectangular permite el cálculo de la distancia mediante el teorema de Pitágoras. Sobre el tiempo, Newton afirma que *"el tiempo absoluto, verdadero y matemático, sin relación a algo exterior, discurre uniformemente y se llama duración"* (Principia, Definiciones [Escolio], 1687).

### La adición de velocidades

La creencia en distancias y tiempos invariantes tiene una consecuencia inmediata en la forma en que se componen dos movimientos. Consideremos dos vehículos, un automóvil y un camión, que avanzan con sentidos opuestos sobre una carretera. Las velocidades del automóvil y el camión relativas a tierra son 100 km/h y 50 km/h respectivamente. Esto significa que en un laboratorio fijo a tierra se verifica que la distancia entre los móviles se reduce en 150 km cada hora. Este mismo fenómeno puede ser examinado desde un laboratorio instalado sobre el automóvil. Si las distancias y los tiempos son invariantes, entonces en ese laboratorio también se verifica que la distancia se acorta 150 km al paso de cada hora. Por lo tanto, la velocidad del camión en el laboratorio del automóvil es de 150 km/h. Lejos de ser invariantes, las velocidades cambian cuando se las examina desde laboratorios con distintos estados de movimiento. Pero, si las distancias y los tiempos son invariantes, ese cambio es muy simple pues se reduce a una adición de velocidades. Aunque el caso considerado es elemental –los movimientos son paralelos y enfrentados–, la regla es válida también en otras situaciones, pero con una adición de velocidades de carácter vectorial.

De la adición de velocidades resulta una propiedad igualmente notable. Supongamos que la velocidad del camión relativa a tierra varía en el tiempo, mientras que la del auto se mantiene constante. Digamos que el camión acelera (cambia su velocidad) a razón de 1km/h cada segundo en el laboratorio fijo a tierra. Pues bien, desde el laboratorio instalado en el auto se observará la misma variación. Esto es así porque las velocidades del camión medidas en cada laboratorio deben diferir siempre en la magnitud de 100 km/h (que es la velocidad constante del auto respecto de tierra). Por lo tanto, cualquier variación de la velocidad del camión respecto de tierra se registra con el mismo valor en el laboratorio del auto. En conclusión, la velocidad de un móvil no es invariante; pero sí lo es la aceleración de un móvil respecto de laboratorios que no experimenten aceleración relativa (como es el caso del auto y la tierra).

### El Principio de Relatividad y las leyes de la Mecánica

En los Principia, Newton adopta como punto de partida el concepto moderno de inercia enunciado por Galileo (1564-1642). Galileo descubre que, contra todas las apariencias, lo esencial del movimiento es su tendencia a perdurar. Newton eleva esta idea a la categoría de Principio, afirmando que un cuerpo libre de fuerzas persiste en un estado de movimiento rectilíneo uniforme (*Principio de inercia* o *Primera Ley de la Dinámica*). Por otra parte, si el móvil se aparta del movimiento rectilíneo y uniforme, Newton afirma que su cambio de velocidad –la aceleración– es proporcional a la fuerza que se ejerce sobre el mismo. Esta es la *Segunda Ley de la Dinámica*, que se enuncia como **F** = *m* **a**. La constante de proporcionalidad *m* entre la fuerza y la aceleración es una propiedad del móvil llamada masa, que expresa la magnitud de su *inercia* (su resistencia a cambiar de estado de movimiento).

Un problema evidente de estas leyes es que no pueden enunciarse para cualquier laboratorio. El movimiento rectilíneo y uniforme del cuerpo libre de fuerzas, ¿respecto de qué laboratorio está considerado? Un movimiento rectilíneo y uniforme es un movimiento con aceleración nula; como fue explicado en la sección anterior, la aceleración mantiene su valor ante cambio de laboratorio sólo si los laboratorios considerados no experimenta entre ellos aceleraciones relativas. De manera que sólo tiene sentido afirmar que un movimiento es rectilíneo y uniforme en alguna familia de laboratorios sin aceleración relativa, pero no en cualquier laboratorio. Los laboratorios donde se puede utilizar el Principio de inercia se llaman laboratorios *inerciales*. ¿Cuál es esta familia de laboratorios privilegiados donde se cumple el Principio de inercia? Podría pensarse que esta pregunta se responde mediante un experimento: bastaría con observar el movimiento de un cuerpo libre de fuerzas para establecer la familia de laboratorios inerciales de una vez y para siempre. Pero, ¿podemos estar seguros de que un cuerpo está completamente libre de fuerzas? En realidad no, a menos que exista un solo cuerpo en todo el espacio. Por lo tanto, la vía del experimento únicamente puede conducirnos al conocimiento de laboratorios que son aproximadamente inerciales. En el marco conceptual de la Física newtoniana, el espacio absoluto –ese escenario abstracto e indefinido– es quien otorga el privilegio a los laboratorios inerciales: un laboratorio inercial es aquel que se traslada uniformemente respecto del espacio absoluto.

Los laboratorios aproximadamente inerciales de nuestra vida cotidiana, además de la tierra, son los medios de transporte como aviones, trenes, barcos (en tanto no estén sometidos a cambios violentos de estado de movimiento como frenadas, trayectorias curvas, etc.). En esos laboratorios no sólo se cumple el Principio de inercia en forma más o menos aproximada, sino cualquiera de las leyes fundamentales de la Mecánica, sin que importen los movimientos relativos uniformes entre ellos. Si arrojamos un objeto dentro de un avión en vuelo, el objeto trazará respecto del avión una trayectoria que será idéntica a la que se obtendría en tierra si lo arrojásemos de la misma manera. Tanto el avión como el laboratorio terrestre son igualmente buenos para verificar las leyes fundamentales de la Mecánica. Estos dos laboratorios están en pie de igualdad. No hay uno mejor que el otro. No hay uno quieto y otro en movimiento, porque no hay ningún fenómeno mecánico que permita dirimir esa cuestión. Lo único que hay entre el avión y la tierra es un movimiento relativo. ¿Quién no ha experimentado la extraña sensación que se tiene a bordo de un tren cuando se observa otro tren contiguo en movimiento relativo, sin que se pueda establecer cuál de los dos se mueve respecto de la estación? Esta equivalencia de los laboratorios inerciales en cuanto a la validez de las leyes fundamentales de la Mecánica se conoce con el nombre de Principio de relatividad.

La cuestión del Principio de relatividad y las leyes de la Mecánica puede presentarse más académicamente, atendiendo a que, por un lado, la aceleración es invariante (como ya vimos). Por otro lado, las fuerzas fundamentales de la Física newtoniana, como la fuerza gravitatoria, dependen de las distancias entre las partículas interactuantes, y por lo tanto son también invariantes. Así, ambos miembros de la *Segunda Ley de la Dinámica* son invariantes (la masa, como propiedad de un cuerpo, es considerada invariante). Por lo tanto, si la *Segunda Ley de la Dinámica* se cumple en un laboratorio inercial, entonces se cumple en todos ellos. Como vemos, el Principio de relatividad en la Mecánica newtoniana está firmemente enraizado en la invariancia de distancias y tiempos.

•

Una vez, en uno de sus frecuentes viajes entre Rosario y Buenos Aires, Mario se preguntó si acaso no sería posible que las propiedades del espacio y el tiempo fueran distintas de las normalmente atribuidas. Muchas veces las cosas no son como parecen. La superficie de la Tierra parece plana cuando estamos sobre ella. El agua parece una sustancia continua y su estructura molecular no se revela a nuestros ojos. El vidrio parece sólido pues su fluidez sólo se vuelve aparente al cabo de muchos años. En realidad, pensó Mario, estas cuestiones dependen de las escalas con que los fenómenos son observados. Así, la curvatura de la superficie terrestre se hará evidente si observamos nuestro planeta desde una distancia comparable a su tamaño. Podría ocurrir que la aparente invariancia de distancias y tiempos ante cambios de estado de movimiento se debiera nada más a que usualmente observamos estos fenómenos en un rango muy limitado de velocidades relativas. ¿Cuál sería, entonces, la escala de velocidades capaz de revelar un comportamiento diferente de distancias y tiempos?

Jugando con esta posibilidad, Mario realizó un experimento pensado. La cuestión se debería dirimir mediante la cinemática más elemental. Entonces Mario consideró una bolita moviéndose a lo largo de una barra. La velocidad relativa barra-bolita $V$ caracteriza completamente el movimiento relativo examinado. Ahora bien, podemos utilizar distintos laboratorios para estudiar este movimiento relativo. Especulando, Mario imaginó que las longitudes de la barra podrían ser diferentes en cada laboratorio. Para fijar ideas, llamó longitud propia $L_o$ a la longitud de la barra medida en un laboratorio donde la barra se encuentra en reposo. En cambio, en un laboratorio donde la barra se moviese con velocidad $v$, su longitud podría ser diferente, digamos $L_v$. Enseguida Mario notó que en su experimento pensado había dos laboratorios especialmente convenientes: por un lado, claro está, el laboratorio donde la barra está en reposo y mide $L_o$. En este laboratorio la bolita se mueve con velocidad $V$. En este laboratorio, la velocidad $V$ no es más que el cociente entre la longitud $L_o$ y el tiempo $\Delta t$ que la bolita emplea para recorrer la longitud de la barra. Por otro lado, en el laboratorio donde la bolita está en reposo el movimiento relativo se debe a que la barra se mueve con velocidad $V$. En este otro laboratorio, la longitud de la barra es $L_V$, de acuerdo con la especulación en juego. A Mario se le ocurrió que el lapso de tiempo entre los eventos considerados –los pasos de la bolita ante cada extremo de la barra–, también podría ser afectado por el cambio de laboratorio. Mario pensó que, desde el punto de vista del lapso de tiempo entre dos eventos, el laboratorio donde la bolita está en reposo es muy particular, porque ambos eventos ocurren en el mismo lugar (pues la bolita no se mueve). Mario resolvió designar con un nombre especial al lapso de tiempo correspondiente al laboratorio donde los eventos ocurren en la misma posición, y lo llamó *tiempo propio* $\Delta\tau$ (en particular, la duración de su viaje medida con su reloj es un tiempo propio). En el primer laboratorio –aquel donde la barra está fija– el lapso de tiempo sería $\Delta t_V$, dejando así abierta la posibilidad de un tiempo distinto en cada laboratorio. Una vez adoptadas estas definiciones, Mario razonó que en su segundo laboratorio, el tiempo $\Delta\tau$ transcurre mientras la barra realiza un desplazamiento igual a su longitud $L_V$. Entonces la velocidad $V$ de la barra no es más que el cociente entre $L_V$ y $\Delta\tau$. En su experimento pensado, Mario había concluido que la velocidad $V$ del movimiento relativo barra-bolita podía obtenerse de dos maneras diferentes: $V = L_o/\Delta t_V$ y $V = L_V/\Delta\tau$. En suma, $L_o/\Delta t_V = L_V/\Delta\tau$, o

$$\frac{L_V}{L_O} = \frac{\Delta\tau}{\Delta t_V} \qquad (1)$$

¿Qué significaba esta relación? Por cierto que esta ecuación se satisface trivialmente si aceptamos la visión ingenua del espacio y el tiempo: con longitudes y tiempos invariantes la ecuación sólo dice $1 = 1$. ¿Podría decir algo más este resultado de la cinemática elemental? En principio dice que los comportamientos de longitudes y tiempos están íntimamente relacionados: si uno de ellos es invariante, ambos lo son. Pero si uno de ellos dependiese de *V* necesariamente el otro también. El viaje había terminado, y Mario debió abandonar la cuestión.

En su siguiente viaje, Mario recordó su reflexión acerca de las escalas típicas de los fenómenos. Si las distancias y tiempos en verdad dependieran del estado de movimiento de los cuerpos o los laboratorios, quizás fuese necesaria una velocidad de una escala muy distinta a la cotidiana para evidenciar tal fenómeno. Buscando una velocidad muy grande, inexorablemente Mario dio con la velocidad de la luz. La luz se propaga a una velocidad tan extraordinaria que Descartes la consideraba infinita, y Galileo no pudo medirla. Finalmente Roemer (1676) se valió de observaciones astronómicas para obtener su valor aproximado. Había algo seductor en este valor finito de la luz (Mario recordaba un sueño juvenil en donde perseguía un rayo de luz sin alcanzarlo, mientras se preguntaba si podría ver el pasado cuando lo lograra). Claro que Mario no pretendía hacer mover una barra o una bolita a una velocidad comparable a la de la luz, porque nunca había visto tal cosa. Pero consideró que las propiedades del espacio y el tiempo podrían someterse a prueba mediante una composición de movimientos donde intervenga una velocidad tan grande, tan diferente, como es la velocidad de la luz. En efecto, el teorema de adición de velocidades fallaría si longitudes y tiempos no fuesen invariantes. Concretamente, si en el experimento de los dos vehículos reemplazamos al camión por la luz que emiten sus faros, ¿se cumplirá el teorema de adición de velocidades para la composición de la velocidad del auto con la de la luz del camión? Es decir, la suma de las velocidades del auto y la de la luz del camión, ambas medidas sobre la tierra, ¿será igual a la velocidad de la luz del camión medida desde el auto? Mario se decepcionó al notar que la velocidad de la luz es tan grande que para detectar su suma con la comparativamente despreciable velocidad del auto sería necesario un experimento sumamente refinado. No obstante, un experimento de ese tipo tal vez detectara una desviación de la mera adición de velocidades, señalando así que las propiedades del espacio y el tiempo serían distintas a las supuestas por la Física Clásica. Mario consideró que había alcanzado un estadio importante en la formulación del problema. Esto era más que suficiente para una sola jornada, y se entregó al descanso.

### El Principio de Relatividad y las leyes de Maxwell. El éter

Existen fenómenos físicos que privilegian un laboratorio inercial sobre los demás, no porque ese laboratorio tenga algo intrínsecamente diferente al resto, sino porque la naturaleza del fenómeno lo selecciona. Este es el caso de la propagación de ondas mecánicas. Las ondas mecánicas, como las que se producen en la superficie del agua de un estanque, o las vibraciones del aire u otro medio que constituyen el sonido, etc. son descriptas por *ecuaciones de onda* cuyas soluciones sólo sirven en el laboratorio donde el soporte material de la onda (el agua, el aire, etc.) se encuentra en reposo. No debe verse en esto ninguna violación al Principio de relatividad, pues el privilegio de un laboratorio sobre el resto obedece a razones físicas concretas. Por otra parte, los laboratorios inerciales no dejan de estar en pie de igualdad por ello, pues en *cualquier* laboratorio donde el soporte material de la onda se encuentre en reposo serán aplicables las soluciones de la ecuación de onda. Un aspecto característico de la forma de la ecuación de onda es que la velocidad de propagación de la onda aparece escrita en la propia ecuación. Esta velocidad de propagación es una propiedad del medio material donde la onda se propaga. Es evidente que una ley física que contenga una velocidad no puede ser válida en cualquier laboratorio, porque las velocidades –contrariamente a las aceleraciones y las distancias– se modifican al cambiar de laboratorio. En otras palabras, la ecuación de onda no podría ser válida en dos laboratorios en movimiento relativo, porque si así fuese la velocidad de la onda resultaría igual en ambos laboratorios. Pero sabemos que las nociones clásicas de espacio y tiempo prohíben que la velocidad de un móvil sea igual en dos laboratorios en movimiento relativo.

En la segunda mitad del siglo XIX, James Clerk Maxwell (1831-1879) dio su forma definitiva a las leyes del electromagnetismo, amalgamando las ya conocidas leyes electrostáticas y magnetostáticas de tal manera que ambos campos –el eléctrico y el magnético– aparecieron como aspectos de una única entidad electromagnética. Esa nueva teoría conducía a ecuaciones de onda para el campo electromagnético. La velocidad de propagación $c$ de esas ondas resultó ser muy semejante a la ya conocida velocidad de la luz, lo que llevó a Maxwell a comprender que la luz es una onda electromagnética. El modelo ondulatorio de la luz, –propuesto por Huygens en 1678 y acabado por Fresnel hacia 1823– había ya prevalecido sobre el modelo corpuscular desde que Foucault comprobara en 1849 que la propagación de la luz en agua es más lenta que en aire, dando la razón al modelo ondulatorio. El modelo ondulatorio de la luz, ahora con status electromagnético, precisaba un sustento material, pues todos los fenómenos físicos eran concebidos de un modo mecanicista. En esto coincidía toda la comunidad científica, incluido el propio Maxwell. Por lo tanto las leyes de Maxwell sólo eran consideradas válidas en un laboratorio donde ese medio material, el *éter*, se encontrase en reposo. El éter debía llenar todo el espacio, pues estaba presente en todo lugar donde un fenómeno luminoso fuese posible. Este éter universal omnipresente podría considerarse en reposo respecto del espacio absoluto de Newton, salvo por fluctuaciones locales provocadas por las interacciones con otros cuerpos. De alguna manera la presencia del éter materializaba el espacio absoluto de Newton. A pesar de la creencia en la existencia de esta sustancia, resultaba evidente que su detección era harto difícil (una circunstancia bien distinta de lo que ocurre con los medios materiales donde se propagan los otros tipos de onda ya mencionados). Si la detección directa del éter era poco menos que imposible, en cambio se podría medir el estado de movimiento de un laboratorio relativo al éter. Se podría establecer la velocidad de la Tierra respecto del éter universal. Bastaría medir la velocidad de la luz en el laboratorio terrestre; esta velocidad relativa resultaría de la adición de velocidades entre $c$ –la conocida velocidad "absoluta" de la luz– y la aún desconocida velocidad "absoluta" de la Tierra. Por lo tanto la medición de la primera conduciría al conocimiento de la última. Se hicieron muchos experimentos dirigidos a revelar los efectos de la adición de velocidades. El primero fue el de Arago en 1810, pero el más famoso fue el de Michelson-Morley en 1887. Este último fue un delicado experimento de fina interferometría donde un rayo de luz era separado en dos rayos que viajaban en distintas direcciones para luego reunirse formando una figura de interferencia. La figura así formada depende de los tiempos de viaje de cada rayo, y estos, a su vez, de las velocidades de cada rayo relativas al laboratorio. La velocidad relativa de cada rayo es el resultado de la adición de velocidades, que en su forma vectorial más general es sensible a las direcciones de los rayos. Cambiando las direcciones de los rayos (esto se consigue rotando la mesa sobre la cual se realiza el experimento) cambiarán las velocidades relativas y se alterará la figura de interferencia. Sin embargo ningún experimento fue capaz de detectar la adición de velocidades. Ningún experimento fue capaz de revelar el movimiento del laboratorio (la Tierra) respecto del éter. Múltiples argumentos se ensayaron para explicar estos fracasos, todos ellos basados en complejas interacciones entre el éter y los cuerpos en movimiento (arrastre parcial del éter, contracción de FitzGerald-Lorentz, etc.). Pero en 1905 Albert Einstein (1879-1955) sostuvo que el éter no existe.

### El Principio de Relatividad de Einstein

Aunque aparentemente inocente, la idea de abolir el éter tenía consecuencias muy severas que sólo una mente como la de Einstein estaba en condiciones de afrontar. Si el éter no existe, ¿en cuál laboratorio inercial son válidas las leyes de Maxwell del electromagnetismo? Einstein sostuvo que las leyes de Maxwell eran válidas en todos los laboratorios inerciales; de esta forma Einstein despojaba a las ondas electromagnéticas del carácter mecánico que se les atribuía. Si las leyes de Maxwell son válidas en todo laboratorio inercial, entonces la velocidad de la luz tiene el mismo valor $c$ en cualquiera de esos laboratorios; esto explica los resultados nulos de los experimentos dedicados a detectar efectos provenientes de velocidades relativas diferentes de $c$. Einstein incorpora, entonces, las leyes de Maxwell al conjunto de leyes fundamentales que satisfacen el Principio de relatividad, lo que permite bautizar su teoría como Teoría de la Relatividad. La facultad del éter universal de privilegiar un laboratorio ha quedado abolida; el concepto de movimiento absoluto ha quedado vacío de contenido empírico. Sólo los movimientos relativos son detectables. Todavía este paso puede parecer trivial si no se examinan con más cuidado sus revolucionarias consecuencias.

¿Cuáles son las consecuencias revolucionarias de estos cambios? Si la luz viaja con la misma velocidad $c$ en cualquier laboratorio inercial –si la velocidad de la luz es invariante– entonces algo malo hay en la adición de velocidades. En efecto, en el ejemplo del automóvil y el camión sería equivocado decir que la velocidad de la luz de los faros del camión es, por ejemplo, $c$ respecto de tierra, pero $c + 100$ km/h respecto del automóvil. Sin embargo, el teorema de adición de velocidades se basa en nuestra creencia en distancias y tiempos invariantes, una creencia muy sólidamente enraizada en nuestra experiencia cotidiana. ¿Será que nuestra experiencia cotidiana nos engaña, debido a su rango muy limitado de movimientos relativos? Así lo consideró Einstein, quien estuvo dispuesto a abandonar las nociones clásicas de espacio y tiempo para sustituirlas por nuevas nociones que se subordinaran a la invariancia de la velocidad de la luz. Las longitudes y los tiempos no son invariantes en Relatividad, sino que cambian al cambiar de laboratorio inercial. Y lo hacen de manera adecuada para que la velocidad de la luz sea invariante. La mera adición de velocidades no es más que una buena aproximación para velocidades pequeñas, como las del automóvil y el camión (pequeñas comparadas con $c$). El valor de $c$ nos da la escala de velocidades en donde se revelan los comportamientos relativistas de longitudes y tiempos.

Retomando la discusión que condujo a la ecuación (1), no es difícil obtener que la velocidad *c* resulta invariante cuando las longitudes se *contraen* y los tiempos se *dilatan* de acuerdo con las expresiones

$$L_V = L_o \sqrt{1 - \frac{V^2}{c^2}} \qquad \Delta t_V = \frac{\Delta \tau}{\sqrt{1 - \frac{V^2}{c^2}}} \qquad (2)$$

Espacio y tiempo son en realidad aspectos relativos de un único *espacio-tiempo*. En cada laboratorio inercial "recortamos" el espacio-tiempo en espacio y tiempo, obteniendo en cada laboratorio inercial distintos valores para las longitudes y los intervalos de tiempo. Pero no importa cómo hagamos el recorte, siempre la velocidad de la luz tiene el mismo valor. Esta velocidad invariante es el nuevo absoluto espacio-temporal.

El efecto de contracción de longitudes y dilatación de tiempos es tan imperceptible a bajas velocidades, que aun si la velocidad relativa *V* entre dos laboratorios alcanzara el 10 % de *c*, los efectos relativistas sobre longitudes y tiempos serían tan sólo del 0,5 % . No obstante, estos efectos son detectables pues las partículas de altas energías ofician de "laboratorios" que viajan a velocidades muy próximas a *c*. Por ejemplo, se comprueba experimentalmente que el tiempo de decaimiento de una partícula inestable que se mueve con velocidad relativa *V* obedece la dilatación relativista de la ecuación (2).

Aun temo que el lector considere estos descubrimientos sobre la naturaleza del espacio-tiempo como un exagerado refinamiento de una lúcida mente humana, sin mayores consecuencias para nuestra vida cotidiana. En ese caso, invito al lector a recordar que las leyes fundamentales de la Mecánica, y su observancia del Principio de relatividad, se fundan en la invariancia de distancias y tiempos. Por lo tanto la Mecánica debe ser reformulada para armonizarla con las contracciones de longitudes y dilataciones de tiempos. En esa reformulación Einstein encontró que la *conservación de la energía* –uno de las principales derivaciones de la Mecánica– involucraba a la masa del cuerpo en pie de igualdad con las energías propias de la Física newtoniana. Einstein pensó que ese resultado indicaba que la masa era una forma de energía. Según sus estudios, la masa albergaba una cantidad de energía $E = mc^2$. Mientras que en la Física newtoniana se creía en la conservación independiente de la masa, Einstein propuso que la masa podría transformarse en otros tipos de energía, aunque respetando la conservación de la energía total. Einstein sugirió que la energía emitida por las ya entonces descubiertas sales de radio podría deberse a un fenómeno de ese tipo. Las ideas de Einstein sobre la equivalencia masa-energía no tardaron en comprobarse. Hoy en día toda la física experimental de altas energías es una directa verificación del mundo relativista. Por otra parte, las energías liberadas en los procesos de fisión de núcleos pesados, que alimentan la generación eléctrica de las centrales nucleares, y en los procesos de fusión de núcleos livianos, que explican los mecanismos que encienden las estrellas, son el resultado de la conversión de masa en otros tipos de energía.

Esta primera etapa de la Teoría de Einstein recibe el nombre de Relatividad Especial. Luego, en la Relatividad General, Einstein presentará la formulación relativista de la gravitación en un tratamiento que la combina con la cuestión del privilegio de los sistemas inerciales.

En un nuevo viaje Mario se preguntó si ya era hora de cuestionar también el supuesto de la geometría plana de Euclides. Tal vez el espacio-tiempo no sea un mero escenario inmutable donde suceden los fenómenos físicos. Tal vez la geometría del espacio-tiempo participa de los fenómenos, siendo alterada por la materia y la energía distribuidas en él. Las partículas abandonadas a su inercia en una geometría espacio-temporal "curva" tendrán movimientos distintos del movimiento rectilíneo y uniforme dictado por la geometría plana de la Física Clásica y la Relatividad Especial. Pero aún así serán movimientos característicos de la propia geometría e independientes de las propiedades de las partículas. Es curioso que en gravitación también ocurre que los movimientos de las partículas son independientes de sus propiedades. ¿Será entonces posible describir la gravitación como un fenómeno asociado a la geometría del espacio-tiempo? Si la geometría del espacio-tiempo tuviera su propia dinámica, gobernada por leyes específicas, ¿cómo sería la geometría del universo, visto a gran escala? ¿Cambiaría con el paso del tiempo, o se mantendría siempre igual? Una perturbación de la geometría del espacio-tiempo, ¿daría lugar a una "onda gravitatoria" propagándose con velocidad finita? ¿Existirían fenómenos gravitatorios desconocidos en escalas diferentes a la de nuestra experiencia cotidiana? La elaboración de estas ideas requeriría de una larga jornada.